\documentstyle [12pt]{article}
\input epsf
\epsfverbosetrue
\newcommand{\be}{\begin{equation}}
\newcommand{\ee}{\end{equation}}
\newcommand{\bdis}{\begin{displaymath}}
\newcommand{\edis}{\end{displaymath}}

\newcommand{\kb}{{\bf k}}
\newcommand{\ppb}{{\bf p}}
\newcommand{\qb}{{\bf q}}

\begin{document}
\title{On the role of inviscid invariants
in shell models of turbulence.}
\author{ L.~Biferale$^{1}$ and R.~Kerr$^{2}$}
\maketitle
\centerline{$^{1}$  Dipartimento di Fisica, Universit\`{a} ``Tor Vergata''}
\centerline{Via della Ricerca Scientifica 1, I-00133, Rome, Italy.}
\centerline{$^{2}$  Geophysical Turbulence Program}
\centerline{National Center for Atmospheric Research}
\centerline{P.O. Box 3000, Boulder, CO 80307-3000}
\medskip
\begin{abstract}
Non-positive definite, global inviscid invariants
similar to helicity are discussed
for two types of shell models and evidence for a new role
for helicity in Navier-Stokes turbulence is presented.
It is suggested that the extra invariants play the role
of triggering  the intermittent cascade of energy
to small scales characterized by pulses. These invariants also determine
where a transition to chaos appears. New analysis of
numerical experiments with existing models is suggested and a new
class of shell-models where the dynamical interactions
of a second quadratic invariant are closer to those of
helicity in the Navier-Stokes equations is introduced.  The place
of the popular GOY model within this class is discussed.
PACS number 47.27.Eq

 \end{abstract}

\newpage

\section{Introduction}

One of the outstanding questions in our understanding of fully-developed
turbulence is the mechanisms by which the cascade of energy to
small scales is maintained.  That the cascade is intermittent is
well recognized, but the phenomenological and dynamical models
used to address the problem are rarely connected to the dynamics in
the full Navier-Stokes equations.  In this note it is shown that
a popular model for explaining turbulent intermittency, the GOY
model \cite{gledzer,yo}, shares some symmetries
with terms in a decomposition of the spectral Navier-Stokes equations
into the interactions between its helical components \cite{waleffe92}.
The essential common property that is identified in both the GOY
model and in Navier--Stokes is the importance
of interactions between components with oppositely signed helicity.
This new role for helicity supporting an intermittent cascade
contrasts strongly with helicity's previously identified role
in blocking the cascade \cite{andrelesieur77,polifkeshtilman89},
a role that was proposed earlier based on an analogy
to the way magnetic helicity creates force-free states \cite{moffatt69}.

In the GOY model with the standard parameters for three-dimensional
turbulence, interactions between modes with oppositely
signed helicity occur naturally as the sign of helicity
reverses between neighboring shells \cite{leo_lohse_wang_benzi}.
Whether Navier--Stokes turbulence
follows a similar path is a more difficult question because there
are several paths, characterized by interactions between different
components of the helicity, that the cascade can follow.  One way
to investigate this question is to consider several variants of the
GOY models that investigate each path in the full Navier--Stokes
equations individually or in unison.  One question would be how
strongly the statistical behavior of the cascade depends on the
symmetries in the different models.  Another line of investigation
is to determine which path the cascade follows in the
full Navier--Stokes equations.  In this note, preliminary results
following both of these approaches are presented and the direction
of a more complete study is presented.  Also included will be
new analysis of the Kerr-Siggia shell model \cite{kerrsiggia},
which has a cubic,
non-positive definite invariant that shares some dynamical properties
with helicity and served as an inspiration for part of this new proposal.
In all of these cases it will be argued that competition between
the transfer of energy and the transfer of the generalized helicity
could explain the presence of numerically observed chaotic dynamics
and intermittency in the energy cascade.  For the GOY models, new
results on intermittency corrections will be used to illustrate
the importance of the second quadratic invariant in the energy cascade.

In section 2, the new results for the Kerr-Siggia model will be
reported.  In section 3, the GOY model is reviewed and how
the inviscid conserved quantities and the dynamics depend
on the free parameters present in the model is discussed.
In section 4, an argument that predicts the transition from  a trivial
dynamics (dominated by the presence of an attractive fixed point) to
a fully chaotic regime for some critical values of the free parameters
is discussed. Some numerical
results for the energy transfer are also discussed.
In section 5, new versions  of the GOY model are introduced by
considering explicitly the possibility of having shells which
transport positive or negative helicity exactly as occurs in
the Navier--Stokes equations.  Two preliminary calculations
with the full Navier--Stokes equations that support the
importance of the interactions between components with oppositely-signed
helicity are presented.  Some problems that can be studied by using
the new variant of the GOY model are discussed and some new
analysis of the full Navier--Stokes equations that could be done
to illuminate these properties is presented.

\section{Pulse scaling of Kerr-Siggia}
The Kerr-Siggia model\cite{kerrsiggia} is a shell model
 with one complex variable
per shell,
originating from a decimation of the possible interactions between triads
in  Burgers equation.  With
a simple forcing and eddy viscosity the equations were
$$du_1/dt = \epsilon/u_1^* + 2iu_2u_1^*$$
\be
du_n/dt = k_{n-1}i(u_{n-1}^2 + 2u_{n+1}u_n^*)
\label{eq:kseq}
\ee
$$du_N/dt=k_{N-1}i(u_{N-1}^2 + \nu_e|u_N|u_N)$$
where $\epsilon$ was the average energy input and dissipation,
 $\nu_e$ was taken to be $2^{2/3}$ and $k_n=2^n$.
Defining
$$E_n={1\over 2}u_n u_n^* $$
\be
H_n=\Re(u_n^* u_{n-1}^2)
\label{eq:ksconserve}
\ee
$$A_n=\Im(u_n^* u_{n-1}^2)$$
for $\epsilon=\nu_e=0$
there are two inviscid invariants $E=\sum{E_n}$ and $H=\sum{H_n}$.
The first is energy and the second, while not positive-definite,
can be treated as a Hamiltonian with canonical variables
$u_n$ and $u_{-n}=u^*_n$ as follows:
\be
du_n/dt = ik_n\delta H/\delta u_{-n}
\label{eq:kscanon}
\ee
The energy transfer between shells is $\epsilon_n=-k_{n-1}A_n$.
There is a trivial, unstable ``Kolmogorov'' solution
of with $u_n=-i(2^{1/3}\epsilon)^{1/3}k_n^{-1/3}$, $H_n=0$,
and $E_n=2^{2/9}\epsilon^{2/3}k_n^{-2/3}$ corresponding
to a solution of an earlier cascade model\cite{des_novikov74} .

The context for discussing this model along with the GOY model
and Navier-Stokes is the extra invariant $H$.  Despite the fact that this
$H$ is cubic and not quadratic, and that neither Euler nor GOY has
a Hamiltonian of this form, due to the non-positive definite nature of
$H$, it appears to have some of the same qualitative effects upon the cascade
that we speculate helicity is capable of for Navier-Stokes and GOY.
The point is that while additional invariants can block the energy
cascade, as enstrophy does in two dimensions and helicity does
to some degree in Navier-Stokes\cite{andrelesieur77,polifkeshtilman89},
due to the non-positive definite nature of the invariant there is an escape
route where the cascade can find a way around this blockage.  As
will be demonstrated for the Kerr-Siggia model, this can take the form
of pulses.  From the tools used to demonstrate this for the
Kerr-Siggia model, it will then be demonstrated that there is
weak evidence for analogous phenomena in Navier-Stokes.

In the original discussion\cite{kerrsiggia}, two classes of solutions
besides the trivial ``Kolmogorov'' solution
were discussed.  First, stationary solutions for a small number
of shells with no forcing or dissipation ($\epsilon=\nu_e=0$)
and maximal $H$ were discussed.
Second, forced, dissipative solutions were discussed.
Intermittency was found in the time dependent solutions and
the effect of the extra invariant was noted, particularly as it
affected the slope of the energy spectrum, which was
$<E_n>\sim k_n^{-1/2}$ rather than the Kolmogorov solution,
but what effect the stationary
solutions might exert upon the forced, dissipative time-dependent solutions
was not considered.


For the present calculations, $\epsilon=1$ was chosen, which gives
a characteristic timescale of $t=1$.  Using as initial conditions
$u_1=u_2=(1,1)$, $u_n=0$, $n\geq3$, $N=14$, it took until
$t=3.9$ for the effects of initial transients to dissappear.
Then statistics were taken until $t=6.8$.  Figure 1 shows
the spectra of $<E_n>$ and $<H_n>$ for this period
as well as $k_n^{-1/2}$ curves, confirming the results
of the original paper\cite{kerrsiggia}. Details will be discussed after the
evidence for pulses is presented.


Figure 2 shows $E_n$ and $H_n$ spectra
for a series of moderately spaced times and the time development of
$E$, $H$ and dissipation for this time period.  By moderately
spaced in time it is meant that the times shown are not so
closely spaced so as to show continuous development, but are
close enough to show a relationship between pulses of $E_n$
and $H_n$ and intermittent bursts in the dissipation.

The primary event to focus upon is best illustrated in the
$H_n$ spectra.  In this sequence it starts as a pulse of positive
$H_n$ centered on $n=4$ at $t=6.11$.  It is associated with only
one of several bumps in the energy spectrum at this time and is
not associated with the spike in energy dissipation at $t=6.15$.
This spike in energy dissipation is assocated with one of the higher
shell bumps in the energy spectrum and comes from a pulse
at an earlier time of oppositely signed $H_n$ similar to
the pulse to be described.

Following the appearance of the positive peak of $H_n$ in
$n=4$ at $t=6.11$, this peak breaks off from the
lower shells and slowly propagates to larger shell numbers.
The energy peak associated with it moves in tandem.  Spectra of the
transfer rates of $E_n$ and $H_n$ have also peaks that move
with the pulse.  When the effects of the highest shell, where
the dissipation occurs, are felt, the pulse stalls
at $t=6.29$ before the energy in the pulse suddenly dissipates
at $t=6.34$.  The stalling is the probable source of the
bump in the time averaged energy spectra just before the
dissipation regime.  While this bump is on top of a
spectrum less steep than Kolmogorov ($k_n^{-1/2}$ rather than
$k_n^{-2/3}$), it is qualitatively similar to a bump in the
turbulent energy spectra for atmospheric observations\cite{champagneetal77},
spectral closures\cite{andrelesieur77}
and forced calculations of Navier-Stokes turbulence\cite{kerr85}.
For Navier-Stokes the bump is believed to be associated with a
bottleneck effect\cite{LohseGroeling} where the decrease in the slope
of kinetic energy spectrum in the dissipation regime blocks the free-flow
of kinetic energy just at the boundary between the inertial and dissipation
subranges. While this effect probably
plays some role in the appearance of the bump in the Kerr-Siggia model,
examination of figure 2 suggests a strong role for the stationary
solutions associated with the second invariant.  This comes from
noticing that the cubic invariant is nearly
maximal over the shells covered by the bump.


While this pulse is dissipating at $t=6.34$, the next major pulse of negative
$H_n$ is beginning to move into shell 2 and positive $H_n$
for the major pulse following that is developing in shell 0 from
the forcing.  So a succession of $E_n$ and alternately
signed $H_n$ pulses is suggested.  Clearly this is a simplified
picture as there are minor spikes in dissipation between the major
spikes that are associated with weak pulses with small $H_n$ of
no particular sign.  An example of such a weak pulse is the blip
in $H_n$ at $n=8$ for $t=6.35$ and the rapidly moving $E_n$
at this time.  To quantitatively demonstrate the alternation in
sign of the strong pulses, figure 3 is a contour plot in shell and
time separation of correlations between different shells and
times of $E_n$ and $H_n$. These plots are similar
to contour plots of the energy transfer in forced Navier-Stokes
calculations\cite{kerr90} and also \cite{kida_when} and in
meteorology are referred to as Hovm\"uller diagrams.  These are:
\be
<(F_{n+\Delta_n,t+\Delta t}- \overline{F_{n+\Delta_n}})
(F_{n,t}-\overline{F_{n}})>
\label{eq:ksflux}
\ee
where $F_n$ is either $E_n$ or $H_n$.  Positive correlations
are dark, negative are light.  These plots are for $n=1$, the
second shell.  The effect of a single pulse is the first region of
increasing $\Delta_n$ and $\Delta t$ originating at $(0,0)$.
The propagation is linear after the first few shells.  Starting
at about $\Delta t=0.5$ there is another strong dark region in
the $E_n$ correlation and a strong light region in the $H_n$
correlation. This supports the qualitative picture coming from
watching the time development that there are a succession of
pulses of oppositely signed $H_n$.

The appearance of these pulses raises several questions.  First,
what modification of the stationary solutions can propagate as a unit?  Second,
what causes the alternation in sign of $H_n$ of the pulses, is it
the forcing or is it the nonlinear dynamics?  We will not attempt to answer
these questions.  The point we do want to make is that there is
some connection between the alternation in sign of the extra
non-positive definite conserved quantity
that seems to be associated with the appearance of pulses in the
energy cascade and with intermittency in the model.  In calculations
where the extra invariant is suppressed, intermittency dissappears.
These ideas are supported by noting that the mechanism with which
the conserved quantities are pumped in the system and removed from the system
can influence the scaling laws in
the inertial subrange\cite{Jon_Lee80,LevShe95}.
An extreme example is a calculation of the Kerr-Siggia model
with a Newtonian viscosity\cite{Jon_Lee80} where
the extra conserved quantity is suppressed, there is a Kolmogorov spectrum
and no intermittency.  These are subtle questions that would require
more accurate studies.

How can the pulses be related to the spectra in figure 1?
The $<E_n>$ spectrum in figure 1 goes as $2^{-n/2}$.  By
dimensional arguments one might expect that the $<H_n>$ spectrum
would obey $2^{-3n/4}$, but this is not required since
at any given time $<H_n>$ can have either sign.  In fact,
the $<H_n>$ spectrum is less steep than this and also seems to
follow $2^{-n/2}$.  To understand this, imagine that each
pulse is a coherent package of $E_n$ and $H_n$ traversing
the spectrum, spending on average $2^{-n/2}$ time in each
shell.  Then the time averaged spectra $<E_n>$ amd $<H_n>$ will
both have $2^{-n/2}$ spectra.  This is similar to the argument that has
been used to generate a -5/3 spectrum from fluctuations in a strained
Burgers vortex\cite{lundgren82}.  If
$\Delta t$ spent in each band goes as $2^{-n/2}$ as this
suggests, then this would imply
that the bands in the $(\Delta_n,\Delta t)$ plots should approach
zero slope as $\Delta_n$ increases.   There is some tendency
in this direction for small $\Delta_n$ in figure 3, but for
larger $\Delta_n$ when the stalling noted at $t=6.29$
in figure 2 and dissipation effects are important, the bands
are nearly linear.  Again, the time-averaged $<H_n>$ should not
have any particular sign, as evidenced by shell 4 in figure 1, and
their magnitude $|<H_n>|$ should decrease as the averaging
time is lengthened.  This has been verified by using different
time intervals for the time averaging.

\section{The GOY model}

Given this discussion, let us now examine
properties of the Kerr-Siggia model shared by the GOY model \cite{gledzer,yo}.

 The GOY model has a
very rich dynamical behaviour and it
has been the object of many studies in  recent years
 (see \cite{leo_lohse_wang_benzi,jpv,bbp,pbcfv,bllp,gpz}
for some numerical and analytical results). It is the most popular
shell-model for 3 dimensional turbulence because of its intermittent
properties are very close to the corresponding quantities
in Navier-Stokes equations when the parameters of the
nonlinear terms share some properties with the nonlinear term
in the Navier-Stokes equations.
In particular, for zero viscosity and no external
forcing, when the system has the same conservation laws as a 3D flow:
conservation of energy, of helicity and of volume in phase space.

The dynamical equations are as follows:
\be
\frac{d}{dt} u_n =i\, k_n \left(u^{*}_{n+1}u^{*}_{n+2} +
b u^{*}_{n+1}u^{*}_{n-1}
+c u^{*}_{n-1}u^{*}_{n-2} \right)
 -\nu k_n^2 u_n +\delta_{n,n_0}f
\label{eq:shell}
\ee
where $\nu$ is the viscosity and $f$ is a forcing acting
on a large-scale-shell (for example, $n_0=1$) introduced
to obtain a statistically stationary dynamical state.
This model has  interactions only between first and
second-neighbor shells in the Fourier space. The two parameters
$b,c$ in the nonlinear terms are chosen such as to conserve
energy, $E= \sum_n |u_n|^2$,
for any choice of $\lambda$. The most general choice of parameters is:
\be
b= - \frac{\epsilon}{\lambda};\qquad c= -\frac{1-\epsilon}{\lambda^2}
\label{eq:parameters}
\ee
where $\epsilon$ is the second free parameter in the model.

The GOY model also has a second quadratic invariant beside
energy:
\be
H = \sum_n
 \chi(\epsilon)^n k_n^{\alpha(\epsilon,\lambda)} |u_n|^2.
\label{eq:helicity}
\ee
 While energy conservation is forced by the choice (\ref{eq:parameters}),
 the characteristics of the second
invariant, $H$, change by changing the values of $\epsilon$ and $\lambda$.
When $\epsilon < 1$ this second invariant is not positive-definite
($\chi(\epsilon) = -1$), while
if $\epsilon > 1$ it is positive-definite ($\chi(\epsilon) = +1$).
By remembering that the Navier-Stokes equations are characterized by having a
second inviscid invariant that is positive-definite in 2D (enstrophy)
and non-positive definite in 3D (helicity),
the value $\epsilon=1$ can be identified as the border between a shell model
for 2D turbulence ($\epsilon>1$) and a  shell model for 3D turbulence
($\epsilon < 1$). In the following,
the problem of whether shell models like  GOY-model are a good
representation of 2D turbulence\cite{angelo} is not addressed
and only the range ($0<\epsilon<1$)
where the dynamics should reproduce aspects of a 3D turbulent flow will be
considered.

By looking in detail at the structure of the second invariant, only
when $\alpha(\epsilon,\lambda)=1$
does it have physical dimensions coinciding with Navier-Stokes helicity
\cite{leo_lohse_wang_benzi}.
This defines a line in the plane of free parameters where the inviscid
conservation laws of the GOY model are
very similar to the 3D Navier-Stokes equations.
Because this invariant has the  same physical dimensions
as 3D helicity and it is  non-positive, we will denote it as
the {\it GOY-helicity} in the following.
In the last section, a modified version of the
GOY model will be introduced with
a second invariant having more correspondance with fluid helicity.

A necessary point before going on is that
model (\ref{eq:shell}) has two inviscid fixed
points corresponding to the Kolmogorov scaling $|u_n| \sim k_n^{-1/3}$
(constant flux of energy, zero flux of helicity) and to a
fluxless scaling $|u_n| \sim k_n^{-(1+\alpha)/3}$ (constant flux
of helicity, zero flux of energy) \cite{bllp}.

For this study our interest in this model comes
from the presence in the $(\epsilon,\lambda)$
plane  of a region where the static Kolmogorov-like fixed point
is dynamically unstable \cite{bllp}. The dynamics is
fully chaotic and shows
an intermittent cascade of energy toward small scales
\cite{jpv,pbcfv} with a complex (multifractal) structure
of the attractor in the phase-space.  This intermittency
is  quantified by measuring the scaling exponents $\zeta(p)$
for  the structure functions in the inertial range:
\be
S_p(k_n) = <|u_n|^p> \sim k_n^{\zeta(p)}
\ee
Only very recently \cite{leo_lohse_wang_benzi,gpz} has it been realized that
the second quadratic invariant plays a crucial role in the dynamics
of the model. In \cite{leo_lohse_wang_benzi}, it was found
by varying the two
free parameters $(\epsilon, \lambda)$ in the 3D-physically
relevant region ($0< \epsilon <1; \lambda > 1$) that
along the line of constant helicity ($\alpha(\epsilon,\lambda)=1$),
the model has the same intermittent behaviour.
That is, the set of $\zeta(p)$ depends only  on the value
of $\alpha$, giving, for the first time, numerical evidence
that the dynamics of the model is strongly dependent
on the presence of the second inviscid-invariant. Furthermore,
it has been shown\cite{gpz} that by modifying the nonlinear term
such as to destroy the presence of the second invariant
(but still preserving the inviscid energy conservation) the intermittent
corrections to K41 seem to weaken.

\section{The transition to chaos}

It has been shown\cite{bllp} that by fixing $\lambda$
and varying the $\epsilon$ parameters (and therefore, by changing
$\alpha$)  the GOY model undergoes
a transition to chaos following a ``Ruelle-Takens'' scenario.
In particular, there exists a critical value, $\epsilon_c$
such that for $\epsilon < \epsilon_c$ the Kolmogorov fixed
point $u_n \sim k_n^{-1/3}$ is dynamically stable. Extending this analysis
by changing the ratio between shells in the range $1< \lambda <3$
it is found that the ``Ruelle-Takens'' transition
is quite general\cite{skl}: there
exists a line in the plane ($\epsilon,\lambda$)
which  divides the region where a Kolmogorov-like fixed point
is dynamically stable from a region where the dynamics is chaotic and
intermittent. This qualitative trend of the
transition line appears to beto be  universal, even if the exact
location can be slightly influenced by the forcing and
by the value of viscosity. In the following, a very
simple argument is presented based on the presence of the second non-positive
invariant that predicts, with good  accuracy, the existence
of the transition and its location  for any value
in the plane ($\epsilon,\lambda$).

Consider the two inviscid quadratic invariants,
the energy and the generalized-helicity, and their currents.
First, for zero viscosity and zero forcing energy conservation gives:
\be
\frac{d}{dt}|u_n|^2 = J_{n-1} -J_{n},
\label{eq:flux_en}
\ee
where the energy current $J_n$
\be
J_n = \Im[-\Delta_{n+1} -(1-\epsilon) \Delta_n].
\label{eq:current_en}
\ee
is defined in terms of triple correlations:
\be
\Delta_n = k_{n-1}u_{n-1}u_{n}u_{n+1}.
\ee
The second conservation law for helicity takes the form:
\be
\frac{d}{dt}(-)^{n}k_n|u_n|^2 = L_{n-1}- L_{n}
\label{eq:flux_he}
\ee
where the current of helicity, $L_n$,
from the $n$th shell to the $n+1$th shell is:
\be
L_n = (-)^{n} k_n \Im[\Delta_{n} -\Delta_{n+1}].
\label{eq:current_he}
\ee
Let us suppose, for the moment, that there exists only
one conserved quantity: energy. Then, very standard arguments
\cite{orszag}
tell us that if viscosity is zero, the system tends to equipartition,
corresponding in the GOY model to $|u_n|^2 = const.$. If
one switches on viscous effects, and starts with an initial
configuration with energy concentrated in the first shells, an energy
cascade toward small scales develops. This energy cascade
has been interpreted as the attempt of the system to reach
new equipartition state\cite{orszag}.
This attempt at restoring equipartition is frustrated by viscous dissipation
at small scales that continuously removes energy
and prevents the small scales from reaching
an equilibrium or quasi-equilibrium state.

The Kolmogorov 1941 cascade, describing a smooth and
constant transfer of energy from large scales to small scales
is another way to rephrase this mechanism. But why does the flow
not follow this picture of relaxing
to a smooth and homogeneous transfer of energy and instead
prefers to use a highly intermittent cascades consisting of
bursts and blockages which are the origins
of the  intermittent corrections to the $\zeta(p)$ exponents? This is
where we are proposing that
the second inviscid quadratic quantity enters into the picture.

It is well known that in 2D turbulence the presence
of a second positive-definite quadratic invariant (enstrophy)
does not allow the energy to cascade forward (toward small scales)
\cite{kraichnan}. This is a general result; {\it either:} the two
conserved quantities must transfer in different directions
in Fourier space {\it or: } only one can transfer to small scales.
For example, in 2D turbulence it is widely believed
that there exists a forward transfer of enstrophy
and a backward transfer of energy (inverse cascade).

In contrast, the presence of a second non-positive definite
quadratic invariant, like helicity in 3D turbulence,
is only a minor constraint on the forward
transfer of energy. Moreover, it is a constraint that, due
to the non-positiveness, can have strong spatial and temporal
fluctuations. If this picture is correct, intermittency in the
3D energy transfer could be the result of a competition between
energy and helicity cascades.  Temporal and/or spatial  intermittency
in the energy flux would be the result of switching between a
net transfer of energy (possible due to cancellation
effects in the helicity flux) and  a depletion in the energy-transfer
due to the presence of a non-zero helicity flux.

How can these phenomenological ideas be checked in the GOY model?
In the GOY model, a smooth and non-intermittent
energy-transfer would correspond to dynamics near the Kolmogorov
manifold $u_n \sim k_n^{-1/3}$. This implies that the energy
flux (\ref{eq:flux_en}) is almost constant in the inertial range and that
the helicity flux (\ref{eq:flux_he})
is almost vanishing. This Kolmogorov
behaviour is obtained when the model has a static
stable fixed point. It is natural, then, to ask if
it is possible to understand the transition from the static
behaviour to chaotic dynamics by invoking the second invariant.
By plugging the Kolmogorov solutions
into the expression for  the generalized-helicity
(\ref{eq:helicity}), we obtain:
\be
H = \sum_n
 \chi(\epsilon)^n k_n^{\alpha(\epsilon,\lambda)-2/3}.
\label{eq:helicity2}
\ee
It is therefore clear that, whether the exponent $(\alpha-{2\over3})$
in (\ref{eq:helicity2}) is
positive or negative determines whether $H$ receives most of its important
contribution from small or large scales, respectively. Therefore,
when $\alpha > 2/3$ the second invariant is concentrated at small
scales and, as in 2D turbulence, prevents a smooth forward
transfer of energy. This is reflected by strong intermittency
and large deviations from Kolmogorov scaling.
But, one can imagine that from time to time that it
is still possible to transfer energy if some cancellation effects
lead to an almost zero $H$-flux. On the other hand, when
$\alpha < 2/3$ energy transfers toward
small scales without having any relevant change in $H$, i.e. the
model relaxes in to a trivial Kolmogorov like fixed point.


Figure 4 shows the numerical results\cite{skl} for the transition
from a static Kolmogorov behaviour to chaotic
dynamics by changing $\epsilon$ and $\lambda$. As predicted,
the transition happens near the critical line defined by:
\be
\alpha(\epsilon_c,\lambda_c) = \frac{2}{3},
 \qquad \lambda_c= (1-\epsilon_c)^{-3/2}
\label{eq:prediction}
\ee
The systematic shift of $5\%$ between the
prediction (\ref{eq:prediction}) and the numerical results is probably
due to viscosity as previously discussed\cite{skl}
We believe that this very simple result is a new important confirmation
that the dynamics of the model is strongly influenced by
the second invariant.


Figure 5 shows decaying numerical experiments, that is zero forcing
and non-zero viscosity with energy initially concentrated
at large scales, that compare the dynamical behaviour of the model in
two characteristic regions: case A [$\epsilon=0.1, \lambda =2$],
where there is smooth energy-transfer regime
(on the right side of the critical curve in fig. 4) and case B
[$\epsilon=0.5, \lambda =2$], where the dynamics is
chaotic and intermittent (left side of the curve in fig. 4).
What is interesting is that for case A, when the second invariant does not
introduce any constraint in the energy transfer ($\alpha < 2/3$),
the energy dissipation is a smooth function. This means
that energy is transferred through the inertial subrange
without any blocking and with a power-law behaviour
in time.  But, in case B ($\alpha > 2/3$),
the energy dissipation has a staircase shape in time indicating long periods
when little energy reaches small scales (blocking) interrupted by
short bursts of dissipation. This would be consistent with the suggested role
for helicity where in addition to blocking the transfer,
cancellation in the helicity transfer is associated with
strong dissipation events.


Figure 6 shows the time evolution of the energy current $J_{n}$ and the
helicity current $L_{n}$ through a  shell in the inertial range
where the model is chaotic. Obviously,
the two fluxes are correlated, but the interesting fact is that when
there is a burst of energy, the helicity flux has a
sinusoidal shape, i.e.  a net forward transfer of energy
is only possible if the net averaged transfer of helicity is zero.
Expanding upon the earlier proposal, this suggests blocking of energy transfer
due to competition with helicity, interrupted by strong dissipation
events made possible by brief, intermittent periods of large helicity
fluctuations, but no net helicity flux. The dissipation events
might then be associated with a strong dynamical coupling between
modes with oppositely signed helicity, which would permit
large helicity fluctuations, but no net helicity flux.

\section{A new shell model}

As described in the previous section, the structure of
what we call helicity in the GOY model is only partially
consistent with the helicity in the Navier-Stokes equations . Apart
from the observation that it has the right dimensions
and that it is not positive-definite,
there is an asymmetry between odd and even shells that
does not any counterpart in physical flows.   One means of
overcoming this problem is to introduce two dynamical variables
in each shell, one transporting positive helicity $u_n^{+}$ and the
other transporting negative helicity $u_n^{-}$. The next step
is choosing how to couple these terms.  For this, we will use
a complete decomposition of the three dimensional
Navier-Stokes eqs.\cite{waleffe92}
into a basis where the two independent components of the velocity
field at each wavenumber correspond to two pure helical waves.
In such a basis there are 4 possible independent classes
of triads interactions distinguished by the combination of
helicity transported from each one of the three interacting modes.

Let us fix, for simplicity, the three modes $\qb,\ppb,\kb$ such that $|\kb|
 < |\ppb| < |\qb|$,
and call $u_{s_k}(\kb), u_{s_p}(\ppb), u_{s_q}(\qb)$ the three interacting
modes, where $(s_k, s_p, s_q) = (\pm 1,\pm 1,\pm 1)$ refer to the sign of
helicity in each mode.
Then, it is simple to show that each triad can fall into one
of the four following classes:
\begin{enumerate}
\item  $(s_k, s_p, s_q)= (+,+,+)$, or $(-,-,-)$
\item  $(s_k, s_p, s_q)= (+,+,-)$, or $(-,-,+)$
\item  $(s_k, s_p, s_q)= (+,-,-)$, or $(-,+,+)$
\item  $(s_k, s_p, s_q)= (-,+,-)$, or $(+,-,+)$
\end{enumerate}

Following this decomposition one is led naturally to introduce
4 classes of shell
models, each one corresponding to one of the four independent
classes of triad interaction present in Navier-Stokes eqs.

What is quite remarkable is that the original GOY model belongs
to one of this classes (the fourth).  To demonstrate this, we write
the general equation for this class using
positive helicity shells $u_n^+$ and negative helicity shells $u_n^-$.
$$\frac{d}{dt} u^{+}_n =i\, k_n \left(u^{-}_{n+1}u^{+}_{n+2} +
b u^{-}_{n+1}u^{-}_{n-1}
+c u^{-}_{n-1}u^{+}_{n-2} \right)^*
 -\nu k_n^2 u^{+}_n +\delta_{n,n_0}f^{+} $$
\be
\frac{d}{dt} u^{-}_n =i\, k_n \left(u^{+}_{n+1}u^{-}_{n+2} +
b u^{+}_{n+1}u^{+}_{n-1}
+c u^{+}_{n-1}u^{-}_{n-2} \right)^*
 -\nu k_n^2 u^{-}_n +\delta_{n,n_0}f^{-}.
\label{eq:pmshell}
\ee
for which the conserved energy and helicity are given by:
\be
E = \sum_n   |u_n^{+}|^2 + |u_n^{-}|^2
\ee
\be
H  = \sum_n   k_n(|u_n^{+}|^2 - |u_n^{-}|^2)
\ee
exactly as in the Fourier-helicity decomposition of
Navier-Stokes equation \cite{waleffe92}.
By noticing that in the original
GOY model shells $n$ and $n+2$ have the same GOY-helicity, it can be seen
that (\ref{eq:pmshell}) is formed from two masked and uncorrelated
versions of the original GOY model for the dynamical
evolution of the variables $(u_1^+,u_2^-,u_3^+,...,u_{2n-1}^+,u_{2n}^-)$
and $(u_1^-,u_2^+,u_3^-,...,u_{2n-1}^-,u_{2n}^+)$. Therefore
(\ref{eq:pmshell}) has, by definition, the same behaviour
as the previous model.
{}From this, we think it is of primary importance to study in details
the dynamical behaviour of the other three shell models (corresponding
to the classes 1,2,3), allowing helicity to have all the
dynamical interactions found in Navier-Stokes.
Work in this direction is in progress and will
be reported elsewhere \cite{bbkt}.

\section{Navier-Stokes helicity}
Up to this point only ideal shell models with extra
non-positive definite invariants have been considered and how they
might be extended to more closely resemble the helicity interactions
in the Navier-Stokes equations.  We would like to relate these ideas
directly to the Navier-Stokes equations.  As noted, earlier attempts
at understanding the effects of helicity have emphasized its power
to block the cascade\cite{andrelesieur77,polifkeshtilman89}.
But, despite the blocking power of the extra invariant,
the shell model calculations are indicating
that the cascade can proceed through interactions
between shells where the sign of the extra invariant is opposite.
In a full calculation, we would also want to see what the effects of
helicity in physical space are.  For example, there are
low Reynolds number Navier-Stokes calculations of how vortex rings
link and unlink and can generate and destroy
helicity\cite{ArefZad,MelanderHussain}.
Because spectral properties were not
analyzed and because of the low Reynolds numbers of these
simulations, strong conclusions about the effects of helicity
cannot be made from these calculations. But it can be said that
even though  the initial conditions contained
large-scale helicity, small scale structures appeared and production
of helicity from viscous effects was not strongly blocked.

Therefore, initial conditions that contain significant large-scale
helicity, but show more clearly how a cascade is not blocked,
would be desireable.  Whether or not production of small-scale
helicity by anisotropies plays a role, as has been
suggested\cite{FrischSheSulem_AKA87}, will not be our objective.
Simulations of isotropic, homogeneous turbulence
in a periodic box are traditionally initialized with a given spectrum, but
the phases of individual wavenumbers is completely random.  To
test the effects of helicity we propose constraining
the phases of the velocity components such that the
helicity of the Fourier modes is not all of one sign,
unlike the investigations of the blocking
power of helicity\cite{polifkeshtilman89}.
Since shell models are indicating strong effects from extra invariant
fluctuations even when the average value is zero, this suggests
initializing a Navier-Stokes calculation with net zero helicity.

Several tests of this type have been done, all with one
qualitatively similar feature. Helicity when it first appears in
the spectra pops up in two neighboring bands of opposite sign,
then the bands separate.  Other than this qualitatively feature,
the number of tests is not yet sufficiently
large to make definitive statements.  Two cases are shown in figure 7.
Both cases are $64^3$ simulations where only a small number
of modes in wavenumber band 4 were initialized.
Each has the common feature noted, but a rich variety of additional
features as well.


In case A, each mode is initialized with
maximal helicity, but otherwise the phases were chosen randomly
(by hand) and the net helicity was zero.  Very quickly the
helicity picture changes.  From a zero helicity spectrum, soon
helicity of opposite sign appears in shells on opposite sides
of the initial energy shell.  For a short period, time
sequences show that the helicity peak at higher wavenumber
moves to small scales until it dissipates, leaving net
helicity of the sign of the large scale peak.  It is during
this phase that relative dissipation rate (dissipation/energy)
is largest.  Therefore, through dissipation of helicity at small scales,
large-scale helicity is generated.  Once only the large-scale
helicity peak is left, the blocking action of the helicity
at large-scales becomes important and the dissipation rate
is suppressed.

In case D, the helicity of each mode is chosen to be zero by using
free-slip boundary conditions in along central planes in the box.
All the modes except one use the same free-slip symmetry plane, with
the exception imposed to break this symmetry.  From these initial
conditions helicity does not initial grow around the initial
$k=4$ wavenumber band, but around the resonance band at $k=8$.
Note that once again helicity first appears in neighboring bands
with opposite sign.  Then time sequences show that the bands
move towards opposite ends of the spectrum.  Once again, dissipation
is largest when the high wavenumber helicity band moves into
the dissipation band and is annihilated, then decreases when
large-scale helicity of only one sign remains.

To make this more quantitative, more calculations need to
be done and there needs to be analysis of helicity spectra
and transfer properties.  But what of a relationship to the
shell models?  Investigations of shell models with many more
than 2 variables per shell seem to invariably lead to
reductions in intermittency.  Fully developed turbulence when
viewed as a shell model has an infinite number of degrees of
freedom and should in this sense not be intermittent.  But it
is, and this is understood as being due to coherent structures
in physical space, which in Fourier space implies strong phase
correlations.  The phase correlations therefore prune the number
of paths the cascade can take, returning us to simple models
with few paths and strong intermittency, such as those presented here.
Therefore, to get meaningful comparisons between 3D direct
calculations and shell models, there must be some means of
identifying the paths the cascade will follow and calculating
statistics along these paths.  Given the difficulty of attaining
this, let us make some other suggestions.

First, there needs to be further work on bi-dimensional correlations
of wavenumber and time, similar to figure 3 here and in earlier
analysis\cite{kerr90,kida_when}.
The question with direct calculations is what quantities to use.
It has been found\cite{kerr90} that energy transfer spectra have a strong
signature.  Helicity spectra and helicity transfer spectra
need to be analyzed in the same manner.  For simulations
with a small number of initial modes, such as the examples just
given, at least for short times statistics for modes formed
by the initial interactions and their daughters could be studied.
It is our hope that analysis of new Navier-Stokes simulations of this
type, coupled with any new understanding fo the role of helicity
coming from shell models, will provide new insight into the nature
of the intermittent cascade of energy to small scales in turbulent
flows.

\vskip 0.3cm
{\bf Acknowledgements}
We thank Detlef Lohse, Leo Kadanoff and
Norbert Schoerghofer for having communicated to us their numerical
data before  publication. We thank, also, Roberto Benzi, Uriel Frisch
Detlef Lohse, Leo Kadanoff, Giovanni Paladin,
Norbert Schoerghofer, Elisabetta Trovatore and  Angelo Vulpiani for useful
discussion.

One of us (LB) has been partially supported by
the  EEC Contract ERBCHBICT941034.  NCAR is supported by the National
Science Foundation.

\newpage
FIGURE CAPTIONS

\begin{enumerate}
\item Time-averaged spectra for the kinetic energy (solid line)
and Hamiltonian (dashes) for the Kerr--Siggia model.  Note the
``bump'' in the energy spectra near shell 9.  The gap in
the Hamiltonian spectra at shell 4 is where it is negative.
$2^{-n/2}$ curves are drawn for comparison.
\item Spectra at individual times for the kinetic energy and Hamiltonian.
In order, the times correspond to 1st time: ---, 2nd time: -- --,
and 3rd time: $\cdot\cdot\cdot$.  Also given is the time evolution
of the total kinetic energy ---, Hamiltonian -- -- and
energy dissipation  $\cdot\cdot\cdot$ for the period covered by the spectra.
The {\bf X} in the energy curves indicates roughly the times of the
spectra to the left.
\item Hovm\"uller diagrams in shell and time separation of kinetic energy
and Hamiltonian fluctuations at different shell numbers and times using
(\ref{eq:ksflux}).  Negative fluctuations correlations are light, positive
are dark.  Note the strongest black for energy is not at (0,0), indicating
that the strongest fluctuations in energy are in the higher shells.
The sign of $H$ fluctuations changes between pulses (light upper ridge),
but does not for energy (dark upper ridge).
\item Comparision in the $(\epsilon,\lambda)$ plane of
the, numerically estimated, transition (circles) \cite{skl} and the theoretical
prediction (solid line).
\item  Log-log plot of the total energy as a function of time in a pure
decaying simulation. Case A (dotted line) corresponds to a smooth
Kolmogorov-like transfer of energy, while case B (solid line) corresponds
to a chaotic intermittent energy-transfer.
\item Fluxes of energy and helicity during a burst trough a shell
in the inertial range. Notice the oscillatory behavior of the helicity
flux triggering the energy transfer.
\item Navier--Stokes three--dimensional banded wavespectra of kinetic energy
and helicity at individual times. At $t=0$ only energy is shown for each case.
For case A, helicity at later time ($t=0.75$) is multiplied by 10.  For
case D, helicity at earlier time ($t=0.0625$) is multiplied by 10.
Triangles indicate helicity maxima at the two times for each case
and crosses indicate helicity minima.
\end{enumerate}

\end{document}